\documentclass[english,aps,preprint,showpacs,preprintnumbers,amsmath,amssymb]{revtex4}
\usepackage[latin1]{inputenc}
\usepackage{babel}

\usepackage{array}
\usepackage{float}
\usepackage{amsmath}
\usepackage{graphicx}
\usepackage[unicode=true, pdfusetitle,
 bookmarks=true,bookmarksnumbered=false,bookmarksopen=false,
 breaklinks=false,pdfborder={0 0 1},backref=false,colorlinks=false]
 {hyperref}

\makeatletter

\usepackage{amsfonts}

\usepackage{slashed}

\makeatletter

\date{\today}

\makeatother

\begin{document}

\title{Renormalization Group and Conformal Symmetry Breaking in the Chern-Simons
Theory Coupled to Matter}

\author{A. G. Dias and A. F. Ferrari}

\affiliation{Universidade Federal do ABC, Centro de Ci\^encias Naturais e Humanas,
09210-170, Santo Andr\'e, SP, Brasil}

\email{alex.dias, alysson.ferrari@ufabc.edu.br}
\begin{abstract}
The three-dimensional Abelian Chern-Simons theory coupled to a scalar
and a fermionic field of arbitrary charge is considered in order to
study conformal symmetry breakdown and the effective potential stability.
We present an improved effective potential computation based on\textbf{
}two-loop calculations and the renormalization group equation: the
later allows us to sum up series of terms in the effective potential
where the power of the logarithms are one, two and three units smaller
than the total power of coupling constants (i.e., leading, next-to-leading
and next-to-next-to-leading logarithms). For the sake of this calculation
we determined the beta function of the fermion-fermion-scalar-scalar
interaction and the anomalous dimension of the scalar field. We shown
that the improved effective potential provides a much more precise
determination of the properties of the theory in the broken phase,
compared to the standard effective potential obtained directly from
the loop calculations. This happens because the region of the parameter
space where dynamical symmetry breaking occurs is drastically reduced
by the improvement discussed here.
\end{abstract}


\maketitle
\global\long\def\LL{{\rm LL}}
 \global\long\def\NLL{{\rm NLL}}
 \global\long\def\NNLL{{\rm N2LL}}

\section{Introduction}

Chern-Simons (CS) theory~\cite{DeserJackiwTempleton} is an important
theoretical framework which has been used to study many issues on
quantum field theory in three space-time dimensions. Among the interesting
properties of CS theory are the classical conformal invariance and
the fact that the gauge field does not receive infinite renormalization,
leading to a zero beta function for the gauge coupling constant~\cite{ChenSemenoff}.
These are important aspects for the problem of symmetry breaking through
radiative corrections~\cite{CW}, which we want to revisit in this
work considering a CS theory coupled to matter.

Our study is motivated by some recent developments concerning the
summation of the power series in the leading and subleading logarithm
terms of the effective potential by means of the renormalization group
equation (RGE)~\cite{elias-charac,elias-SM}. The RGE allows one
to obtain extra information from the usual loop approximation, thus
providing more refined information concerning quantum properties of
the model under scrutiny. An important example where the RGE have
dramatically improved the information obtained in the loop approximation
is in the analysis of the effective potential for the Standard Model
with conformal invariance: from the standard one-loop approximation,
the effective action of the model does not seems to be stable, but
with the more precise approximation obtained using the RGE, one discover
it actually is~\cite{elias-SM}. For other examples see~\cite{elias-charac,diasLL,Meissner,Nicolai}.

We show here an improved calculation of the effective potential of
the theory of a CS field coupled to scalar and fermionic fields. The
computation includes infinite summations of terms of the effective
potential which can be carried out with the RGE and the knowledge
of the elements figuring in it: the beta functions, scalar field anomalous
dimension and the first logarithm corrections for the effective potential.
These elements at lowest approximation need a two-loop calculation
to be determined since there are no one-loop divergences in odd space-time
dimensions when using the sort of regularization we adopt here (regularization
by dimensional reduction). In fact all one particle irreducible diagrams
with an odd number of loops will be finite under this scheme. Some
of the needed elements were computed in Refs~\cite{ChenSemenoff,Hosotani,avdeev92,alves2000,diasCS};
in this work, we calculate the fermion anomalous dimension and the
beta function for the Yukawa coupling.

A peculiarity of CS theory has to be mentioned at this point. The
theory involves the Levi-Civita tensor which cannot be easily extended
to arbitrary dimensions as needed in the context of dimensional regularization.
A regularization procedure called dimensional reduction\,\cite{siegel}
has been shown to be appropriated in dealing with CS theory~\cite{ChenSemenoff,avdeev92,alves2000}:
essentially, it consists in performing the tensor and gamma matrices
algebra in three dimensions, and extending only the momentum integrals
to arbitrary dimensions.

The two-loop results, in conjunction with the RGE, allows us to sum
up all terms in the effective potential where the total power of the
coupling constants is one and two units larger than the power of the
logarithms $\log\left(\phi/\mu^{2}\right)$ (called \emph{leading
logarithms}, LL, and \emph{next-to-leading logarithms}, NLL, terms),
as well as some subseries where they are three units larger (the \emph{next-to-next-to-leading
logarithms} terms). We study the dynamical symmetry breaking of the
conformal symmetry in this theory, showing that the improved effective
potential leads to a much finer determination of the properties in
the broken phase, such as mass and coupling constant of the scalar
field. This happens because the region of the parameter space of the
theory, where the dynamical breaking of symmetry is operational at
the perturbative level, is much smaller when considering the improved
effective potential than for the initial two-loop potential. Another
interesting aspect is that, for certain values of the parameters,
we found two broken vacua, which leads to different physical properties.
This happens both for the improved and the original effective potential,
but the region of the parameter space where this happens is much more
restricted for the former case. Again, the improvement of the perturbative
effective potential calculation provides more precise determination
of the properties of the theory.

We believe that the outcomes of our analysis involving the Chern-Simons
theory enforces the idea that one has to extract the maximum amount
of information from a given perturbative calculation, by using the
renormalization group equations to obtain a better approximation to
the effective potential. Even if its natural at a first moment to
use one-loop results to predict masses and coupling constants from
any of the many proposed extensions to the Standard Model, for example,
one should enrich the analysis of the dynamical symmetry breaking
by means of the RGE. 

This paper is organized as follows. The method of using the RGE to
sum up series of perturbative corrections to the effective potential
is outlined in Sec.\,\ref{sec:General-Considerations}. The model
we shall study is described in Sec.\,\ref{sec:The-Model}. Technical
details of the two-loop calculations needed for this work are presented
in Sec.\,\ref{sec:appA}. Sec.\,\ref{sec:AppB} contains the detailed
calculation of the improved effective potential, which is used to
study the dynamical breaking of the conformal symmetry in Sec.\,\ref{sec:Dynamical-Breaking}.
Finally, our conclusions are summarized in Sec.\,\ref{sec:Conclusions}.

\section{\label{sec:General-Considerations}General Considerations}

We start by reviewing the use of the RGE to calculate the improved
effective potential. As discussed in~\cite{elias-charac}, the standard
practice for solving the RGE by replacing the couplings in the effective
potential by their running values amounts to a particular application
of the method of characteristics to solve partial differential equations.
This procedure does not exhaust, however, the information that is
contained in the RGE: actually, a finer approximation can be obtaining
by writing the effective potential as a general power series in the
couplings and logarithms of the scalar field, and using the RGE to
sum up some infinite subsets of this power series. 

To explain the general procedure, we will consider a general model
of a scalar field $\varphi$ with a self-interaction of the form $\varphi^{N}$,
together with interactions with other dynamical fields. As known,
in three spacetime dimensions, renormalizability imposes that $N\leq6$,
but we shall not fix any particular value of $N$ in this Section.
Let $\lambda=\left\{ \lambda_{i},\, i=1,\ldots M\right\} $ denote
collectively the set of all coupling constants of the theory. The
RGE for the regularized effective potential $V_{\mbox{eff}}\left(\phi\right)$
reads \begin{equation}
\left[\mu\frac{\partial}{\partial\mu}+\beta_{\lambda}\frac{\partial}{\partial\lambda}-\gamma_{\varphi}\phi\frac{\partial}{\partial\phi}\right]V_{\mbox{eff}}\left(\phi;\mu,\varepsilon,\lambda,L\right)=0\,\label{eq:RG1}\end{equation}
(in this section, the sum over all $\lambda_{i}$ will always be implicit).
Here, $\mu$ is the arbitrary mass\textbf{ }scale introduced when
we use dimensional regularization to extended the theory to dimension
$D$, $\gamma_{\varphi}$ is the anomalous dimension of the scalar
$\varphi$, $\varepsilon=3-D$, \begin{equation}
L=\ln\frac{\phi^{2}}{\mu}\,,\label{eq:defL}\end{equation}
and $\phi$ is the vacuum expectation value of the scalar field $\varphi$. 

For the sake of convenience, we introduce the notation (from now on,
we omit the explicit dependence on the parameters $\mu,\varepsilon,\lambda,L$),
\begin{equation}
V_{\mbox{eff}}\left(\phi\right)=\phi^{N}S_{\mbox{eff}}\left(\phi\right)\,,\label{eq:defSeff}\end{equation}

\noindent where $S_{\mbox{eff}}\left(\phi\right)$, on very general
grounds, is a sum of terms involving different powers of $\lambda$
and $L$, which in principle can be calculated order by order in the
loop expansion.

In order to use the RGE we shall organize the terms in $S_{\mbox{eff}}\left(\phi\right)$
according to the power of $L$ relative to the aggregate powers of
the couplings $\lambda$, i.e., \begin{equation}
S_{\mbox{eff}}\left(\phi\right)=S_{\mbox{eff}}^{\LL}\left(\phi\right)+S_{\mbox{eff}}^{\NLL}\left(\phi\right)+S_{\mbox{eff}}^{\NNLL}\left(\phi\right)+\cdots\,,\label{eq:expSeff}\end{equation}
 where \begin{equation}
S_{\mbox{eff}}^{\LL}\left(\phi\right)=\sum_{n\geq1}C_{n}^{\LL}\lambda^{n}L^{n-1}\,,\label{eq:SeffLL}\end{equation}
 is the sum of the \emph{leading logarithms} in $S_{\mbox{eff}}\left(\phi\right)$,
and \begin{equation}
S_{\mbox{eff}}^{\NLL}\left(\phi\right)=\sum_{n\geq3}C_{n}^{\NLL}\lambda^{n}L^{n-2}\,,\label{eq:SeffNLL}\end{equation}
 \begin{equation}
S_{\mbox{eff}}^{\NNLL}\left(\phi\right)=\sum_{n\geq3}C_{n}^{\NNLL}\lambda^{n}L^{n-3}\,,\label{eq:SeffN2LL}\end{equation}
are the \emph{next-to-leading} and \emph{next-to-next-to-leading logarithms}
terms, respectively; here, $\lambda^{n}=\prod\lambda_{i}^{n_{i}}$
with $\sum n_{i}=n$. The RGE allows one to calculate these sums once
their first coefficient is known, if we have enough information on
the $\beta$-functions and the anomalous dimension of the scalar field.
To see how this come about, we use the definition (\ref{eq:defSeff})
in Eq.\,(\ref{eq:RG1}), and take Eq. (\ref{eq:defL}) into account
to rewrite the RGE in a more convenient form, \begin{equation}
\left[-\left(1+2\gamma_{\phi}\right)\frac{\partial}{\partial L}+\beta_{\lambda}\frac{\partial}{\partial\lambda}-N\gamma_{\varphi}\right]S_{\mbox{eff}}\left(\phi;\mu,\varepsilon,\lambda,L\right)=0\,.\label{eq:RG2}\end{equation}

\noindent We shall write $\gamma_{\varphi}$ and $\beta_{\lambda}$
in the form \begin{align}
\gamma_{\varphi} & =\gamma_{\varphi}^{\left(2\right)}+\gamma_{\varphi}^{\left(3\right)}+\cdots\,,\\
\beta_{\lambda} & =\beta_{\lambda}^{\left(2\right)}+\beta_{\lambda}^{\left(3\right)}+\cdots\,,\end{align}
where $\gamma_{\varphi}^{\left(j\right)}$ and $\beta_{\lambda}^{\left(j\right)}$
denotes the terms of order $\lambda^{j}$ of the anomalous dimension
and beta function, respectively; these can be obtained by explicit
loop calculations.

Substituting the expansion (\ref{eq:expSeff}) in (\ref{eq:RG2})
we find, at the leading order (terms proportional to $\lambda^{n}L^{n-2}$),
\begin{equation}
\left[-\frac{\partial}{\partial L}+\beta_{\lambda}^{\left(2\right)}\frac{\partial}{\partial\lambda}\right]S_{\mbox{eff}}^{\LL}\left(\phi\right)=0\,.\label{eq:RGELL}\end{equation}
This results in a first order difference equation for the coefficients
$C_{n}^{\LL}$; in this way $S_{\mbox{eff}}^{\LL}\left(\phi\right)$
can be determined once we know $\beta_{\lambda}^{\left(2\right)}$
and the initial coefficient $C_{1}^{\LL}$. Having $S_{\mbox{eff}}^{\LL}\left(\phi\right)$
at our disposal, we can focus at terms of order $\lambda^{n}L^{n-3}$
in (\ref{eq:RG2}), \begin{equation}
\left[-\frac{\partial}{\partial L}+\beta_{\lambda}^{\left(2\right)}\frac{\partial}{\partial\lambda}\right]S_{\mbox{eff}}^{\NLL}\left(\phi\right)+\left[\beta_{\lambda}^{\left(3\right)}\frac{\partial}{\partial\lambda}-N\gamma_{\varphi}^{\left(2\right)}\right]S_{\mbox{eff}}^{\LL}\left(\phi\right)=0\,.\label{eq:RGENLL}\end{equation}

\noindent Since $S_{\mbox{eff}}^{\LL}\left(\phi\right)$ is known,
this equation allows us to calculate $S_{\mbox{eff}}^{\NLL}\left(\phi\right)$
if we have $\beta_{\lambda}^{\left(3\right)}$, $\gamma_{\varphi}^{\left(2\right)}$
and $C_{2}^{\NLL}$.

This procedure can be repeated until we have exhausted the information
on $\beta_{\lambda}$, $\gamma_{\varphi}$ and the initial coefficients
$C$ from the explicit loop calculations. In summary, the RGE allows
one to use the knowledge of $S_{\mbox{eff}}\left(\phi\right)$, $\beta_{\lambda}$
and $\gamma_{\varphi}$ up to a given loop order to sum up complete
subsets of contributions for the effective potential arising from
all loop orders, thus extracting the maximum amount of information
from our perturbative calculation.

\section{\label{sec:The-Model}The Model}

We shall now consider a Chern-Simons field $A_{\mu}$ in three spacetime
dimensions coupled to a two component Dirac field $\psi$ and a complex
scalar field $\varphi$, both charged under the $U\left(1\right)$
gauge symmetry of the CS field according to the Lagrangian\begin{align}
\mathcal{L}= & \frac{1}{2}\epsilon_{\mu\nu\rho}A^{\mu}\partial^{\nu}A^{\rho}+i\overline{\psi}\gamma^{\mu}D_{\mu}\psi+\left(D^{\mu}\varphi\right)^{\dagger}\left(D_{\mu}\varphi\right)\label{eqq:1}\\
 & -\frac{\nu}{6}\left(\varphi^{\dagger}\varphi\right)^{3}-\alpha\varphi^{\dagger}\varphi\overline{\psi}\psi\,.\nonumber \end{align}
The theory has a self-interaction for the scalar field and an Yukawa-like
interaction between scalar and fermions fields. In Eq.\ (\ref{eqq:1}),
$\nu$ is a positive coupling constant and $D^{\mu}=\partial^{\mu}-ienA^{\mu}$,
where $n$ is the charge of the field $D^{\mu}$ is acting on. Without
loss of generality, we can consider $n_{\varphi}=1$, since any $n_{\varphi}\neq1$
can be reabsorbed by a redefinition of the gauge coupling constant
$e$. Therefore, we will denote simply by $n$ the charge of the fermion,
from now on. The spacetime metric is $g^{\mu\nu}=(1,-1,-1)$, the
fully antisymmetric Levi-Civita tensor $\epsilon_{\mu\nu\rho}$ is
normalized as $\epsilon_{012}=1$, and the gamma matrices were chosen
as $\gamma^{\mu}=\left(\sigma^{3},i\sigma^{1},i\sigma^{2}\right)$.

The Lagrangian in Eq.\ (\ref{eqq:1}) is a $\left(2+1\right)$ dimensional
analog of the well known Coleman-Weinberg model in $(3+1)$ dimensions~\cite{CW},
in the sense that all parameters appearing in the classical Lagrangian
are dimensionless, so it posseses classical conformal invariance.
As we assume such an invariance at the classical level, to deal with
quantum corrections it is appropriate to use a regularization method
that violates it minimally~\cite{meissner2007}. The observations
made in~\cite{meissner2007} regarding dimensional regularization
are straightforwardly generalized for regularization by dimensional
reduction, which has been used to obtain the quantities we need here.
Divergent integrals are regulated by the replacement $\int d^{3}k/(2\pi)^{3}\rightarrow\mu^{\epsilon}\int d^{3-\epsilon}k/(2\pi)^{3-\epsilon}$,
where the mass scale $\mu$ is introduced to keep the dimensions of
the relevant quantities unchanged. Conformal invariance is broken
explicitly by this mass scale, but $\mu$ comes with the evanescent
exponent $\epsilon$ and this, in conjunction with the poles $1/\epsilon$,
means that $\mu$ always appears inside a logarithm. Also, regularization
by dimensional reduction has been shown to preserve Ward identities
at least until the two loop order~\cite{alves2000,dias2001}.

Details of the two-loop calculation of the effective potential for
a theory like in eq.\ (\ref{eqq:1}) can be found in~\cite{diasCS}.
In summary, after introducing a convenient gauge fixing, one defines
a Lagrangian $\hat{\mathcal{L}}_{\mbox{int}}$ shifting the scalar
fields by a constant, and disregarding terms independent of or linear
on the fields~\cite{jackiw74}; after that, the effective potential
can be calculated by means of \begin{align}
V_{\mbox{eff}}\left(\phi\right)=\frac{\nu}{48}\phi^{6}-\frac{i}{2}\int\frac{d^{3}\hspace{0.1cm}k}{(2\pi)^{3}}\ln\left[\det\left(i\Delta_{\alpha\beta}^{-1}\left(k,\phi_{i}\right)\right)\right]+i<0|\, T\exp i\!\int d^{3}x\,\hat{\mathcal{L}}_{\mbox{int}}|0>.\label{eq:3}\end{align}

\noindent Hereafter, $\phi^{2}$ stands for $\sqrt{2}\left\langle \varphi^{\dagger}\varphi\right\rangle $.
The first and second terms in Eq.\ (\ref{eq:3}) are, respectively,
the tree approximation and the one-loop correction to the effective
potential; the third term is the sum of the vacuum diagrams with two
and more loops.

We quote here the two-loop effective potential in the following form~\cite{diasCS},
\begin{equation}
V_{\mbox{eff}}^{2\ell}\left(\phi\right)=\frac{\pi}{6}\phi^{6}S_{\mbox{eff}}^{2\ell}\left(\phi\right)\,,\label{eq:5}\end{equation}
 where $S_{\mbox{eff}}^{2\ell}\left(\phi\right)$ is more conveniently
written in terms of the coupling constants \begin{equation}
y=\frac{\nu}{8\pi^{2}}\,\,;\,\, x=\frac{e^{2}}{2\pi}\,\,;\,\, z=\frac{\alpha}{2\pi}\,,\label{eq:6}\end{equation}
 as follows, \begin{align}
S_{\mbox{eff}}^{2\ell}\left(\phi\right)\,=\, y+ & \left[24\left(1+\frac{n_{\psi}^{2}}{8}\right)x^{4}-\frac{33}{2}x^{2}y+14y^{2}+\right.\nonumber \\
 & \left.\frac{15}{4}yz^{2}-3z^{4}-6n_{\psi}^{2}x^{3}z+3n_{\psi}^{2}x^{2}z^{2}\right]L\,.\label{eq:61}\end{align}
On the other hand, as discussed in Section\,\ref{sec:General-Considerations},
the general form for $S_{\mbox{eff}}\left(\phi\right)$ can be cast
as in Eq. (\ref{eq:expSeff}), with \begin{align}
S_{\mbox{eff}}^{\LL}\left(\phi\right) & =\underset{\left(n+m+p\geq1\right)}{\sum_{n,m,p\geq0}}C_{n,m,p}^{\LL}x^{n}y^{m}z^{p}L^{n+m+p-1}\label{eq:8}\end{align}
 \begin{equation}
S_{\mbox{eff}}^{\NLL}\left(\phi\right)=\underset{\left(n+m+p\geq2\right)}{\sum_{n,m,p\geq0}}C_{n,m,p}^{\NLL}x^{n}y^{m}z^{p}L^{n+m-2}\,,\label{eq:9}\end{equation}
 and \begin{equation}
S_{\mbox{eff}}^{\NNLL}\left(\phi\right)=\underset{\left(n+m+p\geq3\right)}{\sum_{n,m,p\geq0}}C_{n,m,p}^{\NNLL}x^{n}y^{m}z^{p}L^{n+m-3}\,.\label{eq:10}\end{equation}

It is known that the beta function of the gauge coupling $x$ vanishes
in CS model coupled to scalar and fermionic fields~\cite{ChenSemenoff};
we calculate the two-loop approximation the beta function $\beta_{\alpha}$
of the Yukawa coupling, as well as the scalar anomalous dimension
$\gamma_{\varphi}$ in Section\,\ref{sec:appA}. The Renormalization
Group equation reads, in our model, \begin{equation}
\left[-\left(1+2\gamma_{\phi}\right)\frac{\partial}{\partial L}+\beta_{y}\frac{\partial}{\partial y}+\beta_{z}\frac{\partial}{\partial z}-6\gamma_{\varphi}\right]S_{\mbox{eff}}\left(\phi\right)=0\,.\label{eq:11}\end{equation}
By following the procedure outlined in Section\ \ref{sec:General-Considerations},
we obtained closed-form expressions for $S_{\mbox{eff}}^{\LL}\left(\phi\right)$,
$S_{\mbox{eff}}^{\NLL}\left(\phi\right)$ and $S_{\mbox{eff}}^{\NNLL}\left(\phi\right)$.
The technical details of this calculation are quite involved and are
developed in Section\,\ref{sec:AppB}. The results we obtain are
the following,\begin{subequations}\label{eq:12} \begin{align}
S_{\mbox{eff}}^{\LL}\left(\phi\right) & =\frac{y}{w}\,,\\
S_{\mbox{eff}}^{\NLL}\left(\phi\right) & =x^{2}S_{\left(2,0\right)}^{\NLL}\left(w\right)+z^{2}S_{\left(0,2\right)}^{\NLL}\left(w\right)\,,\\
S_{\mbox{eff}}^{\NNLL}\left(\phi\right)= & \left(x^{4}S_{\left(4,0\right)}^{\NNLL}\left(w\right)+z^{4}S_{\left(0,4\right)}^{\NNLL}\left(w\right)\right.\nonumber \\
 & \left.+x^{2}z^{2}S_{\left(2,2\right)}^{\NNLL}\left(w\right)+x^{3}zS_{\left(3,1\right)}^{\NNLL}\left(w\right)\right)L\,,\end{align}
\end{subequations} where \begin{equation}
w=1-a_{1}yL\label{eq:13}\end{equation}
and the functions of $w$ appearing in Eq. (\ref{eq:12}) are explicitly
displayed in Section\,\ref{sec:AppB}.

\section{\label{sec:appA}Two-loop wavefunction renormalization and $\beta$
functions}

For the purposes of this work we need to calculate the beta function
for the Yukawa coupling $\alpha\varphi^{\dagger}\varphi\overline{\psi}\psi$,
which implies in calculating the renormalization of the four-point
$\varphi^{\dagger}\varphi\overline{\psi}\psi$ function, as well as
the wave function renormalization of the $\psi$ field. To evaluate
these quantities, we calculated in the two-loop approximation the
divergent parts of the fermion two-point vertex-function $\Gamma_{\overline{\psi}\psi}$
and the four point vertex function $\Gamma_{\varphi^{\dagger}\varphi\overline{\psi}\psi}$.
Free propagators for fermionic, scalar and gauge fields are given
respectively by\begin{subequations}\begin{align}
\Delta_{\psi}\left(k\right) & =\frac{i}{\slashed k-i\eta}\,,\\
\Delta_{\varphi}\left(k\right) & =\frac{i}{k^{2}-i\eta}\,,\\
\Delta_{\mu\nu}\left(k\right) & =\frac{\varepsilon_{\mu\nu\sigma}k^{\sigma}}{k^{2}-i\eta}\,,\end{align}
\end{subequations}while the elementary vertices are\begin{subequations}\begin{eqnarray}
\mbox{trilinear }\overline{\psi}\psi A_{\rho} & \leftrightarrow & -ie\gamma_{\rho}\mu^{\varepsilon/2}\,,\\
\mbox{trilinear }\overline{\varphi}\left(p\right)\varphi\left(-q\right)A_{\rho} & \leftrightarrow & -ie\left(p+q\right)_{\rho}\mu^{\varepsilon/2}\,,\\
\mbox{quadrilinear }\overline{\varphi}\varphi A_{\rho}A_{\sigma} & \leftrightarrow & ie^{2}g_{\rho\sigma}\mu^{\varepsilon}\,,\\
\mbox{quadrilinear }\overline{\psi}\psi\overline{\varphi}\varphi & \leftrightarrow & -i\alpha\mu^{\varepsilon}\,,\end{eqnarray}
\end{subequations}where, in the $\overline{\varphi}\varphi A$ vertex,
the indicated momenta are the ones \emph{entering} the respective
line. 

\begin{figure}[H]
\begin{centering}
\includegraphics[width=0.6\columnwidth]{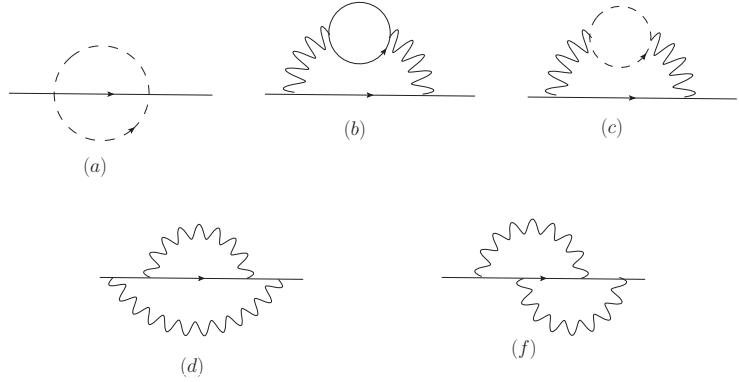}
\par\end{centering}

\caption{\label{fig:fermion-2p}Two-loop contributions to the fermion two-point
vertex function.}

\end{figure}

The diagrams involved in calculating the two-point vertex function
of the fermion are shown in Fig.\,\ref{fig:fermion-2p}, and the
corresponding divergent parts are given by,\begin{subequations}\label{eq:A1}\begin{align}
\left(a\right) & =-\frac{\alpha^{2}}{6}\,,\,\,\left(b\right)=\left(c\right)=\frac{5e^{4}}{36}\,,\\
\left(d\right) & =-\frac{e^{4}}{3}\,,\,\,\left(e\right)=\frac{e^{4}}{2}\,,\end{align}
\end{subequations}apart from an $\left(i\mu^{2\varepsilon}/16\pi^{2}\right)\slashed k/\varepsilon$
factor.

\begin{figure}[H]
\begin{centering}
\includegraphics[width=0.5\columnwidth]{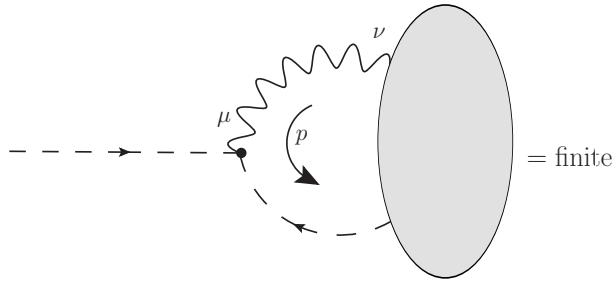}
\par\end{centering}

\caption{\label{fig:theo}A simple rule for establishing the finiteness of
a subset of diagrams: since the external momenta can be taken to zero,
whenever there is a trilinear $\overline{\varphi}\varphi A$ vertex
attached to an external line, the resulting Feynman integrand would
contain a factor $\Delta_{\mu\nu}\times\left(-iep^{\mu}\right)$,
thus vanishing due to the antisymmetry of the gauge propagator $\Delta_{\mu\nu}$.}

\end{figure}

\begin{figure}[H]
\begin{centering}
\includegraphics{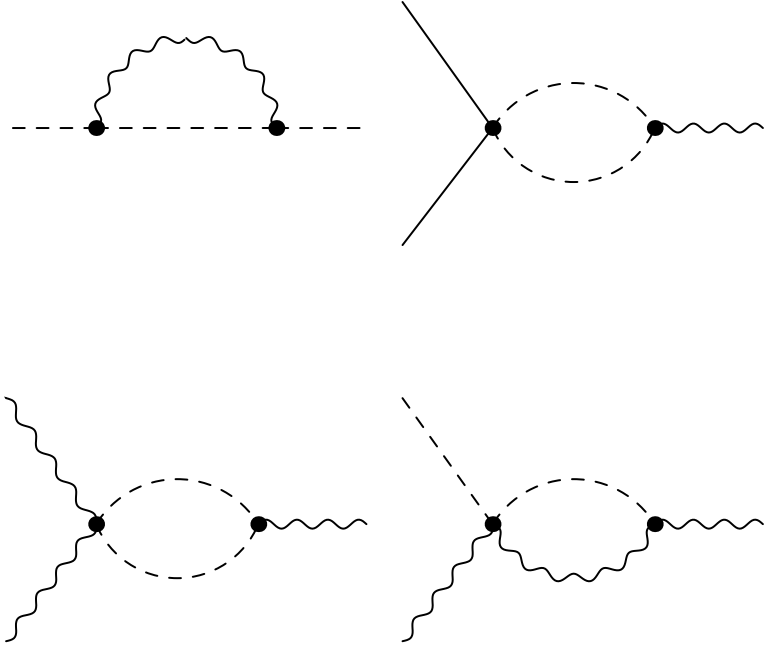}
\par\end{centering}

\caption{\label{fig:vanish}One-loop vanishing diagrams that appear as subgraphs
of some of the two-loop contributions to the four-point vertex function.}

\end{figure}

\begin{figure}[H]
\begin{centering}
\includegraphics{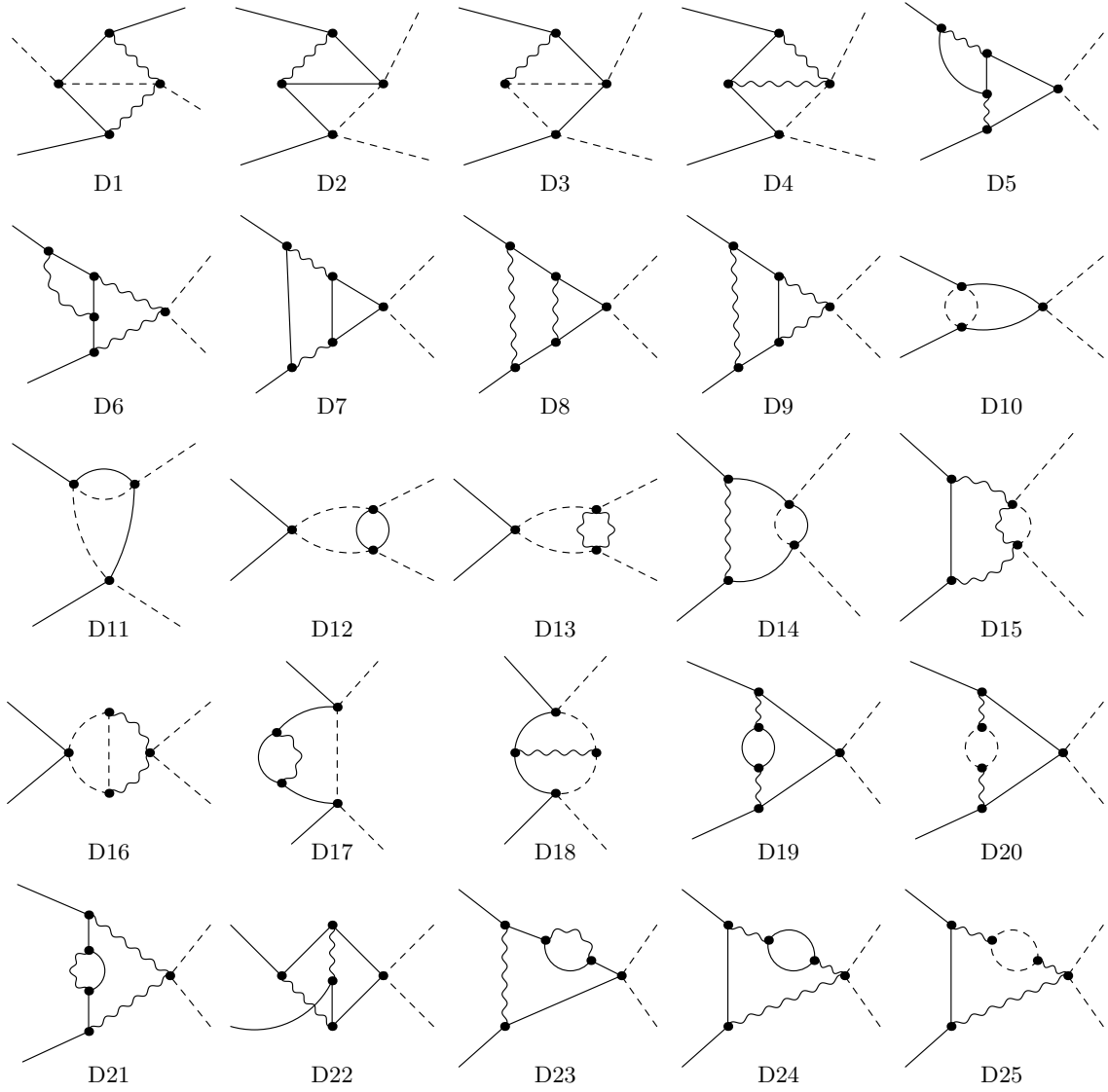}
\par\end{centering}

\caption{\label{fig:diagrams}Potentially divergent two-loop diagrams.}

\end{figure}

We also evaluated the divergent part of the four-point $\overline{\psi}\psi\overline{\varphi}\varphi$
vertex function in the two-loop approximation. Our method for this
calculation was the following one: all two-loop 1PI diagrams for such
vertex function were generated using the \emph{Mathematica} package
FeynArts\,\cite{feynarts}, resulting in about 200 diagrams The
identification of the divergent diagrams was greatly facilitated by
the fact that, for the purpose of evaluating the divergent part of
the $\overline{\psi}\psi\overline{\varphi}\varphi$ function, we could
calculate the diagrams with vanishing external momenta. This allowed
us to prove an important rule, all diagrams with a trilinear $\overline{\varphi}\varphi A$
vertex attached to an external line are finite due to the antisymmetry
of the gauge propagator $\Delta_{\mu\nu}$. This rule is graphically
represented in Fig.\,\ref{fig:theo}. There are also some one-loop
diagrams that vanish (those depicted in Fig. \ref{fig:vanish}) and
appear as subdiagrams of some of the initial set. Using the pattern-matching
capabilities of \emph{Mathematica}, we could use such rules to narrow
down the set of possibly divergent two-loop diagrams to those appearing
in Fig. \ref{fig:diagrams}. The result of the calculation of these
diagrams appears in Table \ref{tab:diagrams}. 

With these results, we can now write down the relation between bare
(denoted by the subscript zero) and renormalized fields and coupling
constants \begin{subequations}\label{eq:A2}\begin{align}
\varphi_{0} & =Z_{\varphi}^{\frac{1}{2}}\varphi=\left(1+A\right)^{\frac{1}{2}}\varphi\,,\\
\psi_{0} & =Z_{\psi}^{\frac{1}{2}}\psi=\left(1+F\right)^{\frac{1}{2}}\psi\,,\\
\alpha_{0}\overline{\varphi}_{0}\varphi_{0}\overline{\psi}_{0}\psi_{0} & =\mu^{\varepsilon}\left(\alpha+\delta\alpha\right)\overline{\varphi}\varphi\overline{\psi}\psi\,.\end{align}
\end{subequations}The constant $Z_{\varphi}$ has already been calculated
in\,\cite{diasCS}, and the results of Eq. (\ref{eq:A1}) and Table\,\ref{tab:diagrams}
allow us to find $Z_{\psi}$ and $\delta\alpha$:\begin{subequations}\label{eq:A3}
\begin{align}
\delta\alpha & =-\frac{1}{32\pi^{2}\varepsilon}\left(7e^{4}\alpha+4e^{2}\alpha^{2}+20e^{6}-4\alpha^{3}\right)\,,\\
Z_{\psi} & =1+\frac{1}{288\pi^{2}\varepsilon}\left(3\alpha^{2}-8e^{8}\right)\,,\\
Z_{\varphi} & =1+\frac{1}{16\pi^{2}\varepsilon}\left[\frac{e^{4}}{3}(7+2n^{2})-\frac{\alpha^{2}}{6}\right]\,.\end{align}
\end{subequations}

The beta function for the Yukawa coupling is calculated from the relation
(\ref{eq:A2}c)\begin{equation}
\alpha_{0}=\frac{\mu^{\varepsilon}\left(\alpha+\delta\alpha\right)}{Z_{\psi}Z_{\varphi}}\,,\label{eq:A4}\end{equation}
since $d\alpha_{0}/d\mu=0$, we have\begin{equation}
\beta_{\alpha}=\mu\frac{d\alpha}{d\mu}=\frac{1}{8\pi^{2}}\left[5e^{6}+\left(\frac{97}{36}+\frac{n^{2}}{3}\right)\alpha e^{4}+\alpha^{2}e^{2}-\alpha^{3}\right]\,.\label{eq:A5}\end{equation}
In terms of the rescaled coupling constants in Eq. (\ref{eq:6}),\begin{equation}
\beta_{z}=\frac{\beta_{\alpha}}{2\pi}=\frac{5}{2}x^{3}+\left(\frac{97}{72}+\frac{n^{2}}{6}\right)zx^{2}+\frac{1}{2}z^{2}x-\frac{1}{2}z^{3}\,.\label{eq:A6}\end{equation}
From Eqs. (\ref{eq:A3}b) and (\ref{eq:A3}c), we obtain the anomalous
dimensions for scalar and fermion fields,\begin{subequations}\label{eq:A7}\begin{align}
\gamma_{\varphi} & =-\frac{1}{Z_{\varphi}}\frac{dZ_{\varphi}}{\mu}=-\left(\frac{7}{12}+\frac{n^{2}}{6}\right)x^{2}+\frac{1}{24}z^{2}\,,\\
\gamma_{\psi} & =-\frac{1}{Z_{\psi}}\frac{dZ_{\psi}}{\mu}=-\frac{1}{9}x^{2}+\frac{1}{24}z^{2}\,,\end{align}
\end{subequations}where $\gamma_{\psi}$ has been quoted just for
completeness.

As for the beta function of the coupling $\nu$, it is most easily
calculated by relating it with the effective potential in Eq.\,(\ref{eq:61})
and the anomalous dimension $\gamma_{\varphi}$ by means of the renormalization
group equation, as done in\,\cite{diasCS}. Here, we just quote the
result, taking into account Eq. (\ref{eq:6}) and the fact that the
fermion has charge $n$, \begin{align}
\beta_{y}= & 24\left(1+\frac{n^{2}}{8}\right)x^{4}-\left(n^{2}+20\right)x^{2}y+14y^{2}\label{eq:A9}\\
 & +4yz^{2}-3z^{4}-6n^{2}x^{3}z+3n^{2}x^{2}z^{2}.\end{align}

\begin{table}[H]
\begin{centering}
\begin{tabular}{|c|c||c|c||c|c||c|c||c|c|}
\hline 
\noalign{\vskip\doublerulesep}
D1 & $-\frac{3}{2}\alpha e^{4}$ & D6 & $-e^{6}$ & D11 & $\frac{1}{2}\alpha^{3}$ & D16 & $-\alpha e^{4}$ & D21 & $-e^{6}$\tabularnewline[\doublerulesep]
\hline 
\noalign{\vskip\doublerulesep}
D2 & $0$ & D7 & $\alpha e^{4}$ & D12 & $\frac{1}{4}\alpha^{3}$ & D17 & $\frac{1}{2}\alpha^{2}e^{2}$ & D22 & $-\frac{1}{4}\alpha e^{4}$\tabularnewline[\doublerulesep]
\hline 
\noalign{\vskip\doublerulesep}
D3 & $-\alpha^{2}e^{2}$ & D8 & $\frac{1}{2}\alpha e^{4}$ & D13 & $-\frac{1}{2}\alpha e^{4}$ & D18 & $0$ & D23 & $-\alpha e^{4}$\tabularnewline[\doublerulesep]
\hline 
\noalign{\vskip\doublerulesep}
D4 & $\alpha e^{4}$ & D9 & $e^{6}$ & D14 & $-\frac{1}{2}\alpha^{2}e^{2}$ & D19 & $-\frac{1}{4}\alpha e^{4}$ & D24 & $-e^{6}$\tabularnewline[\doublerulesep]
\hline 
\noalign{\vskip\doublerulesep}
D5 & $\frac{1}{2}\alpha e^{4}$ & D10 & $\frac{1}{4}\alpha^{3}$ & D15 & $-2e^{6}$ & D20 & $-\frac{1}{4}\alpha e^{4}$ & D25 & $-e^{6}$\tabularnewline[\doublerulesep]
\end{tabular}
\par\end{centering}

\caption{\label{tab:diagrams}Divergent parts of the diagrams appearing in
Fig. \ref{fig:diagrams}, omitting an overall factor of $i\mu^{\varepsilon}/8\pi^{2}\varepsilon$.}

\end{table}

\section{\label{sec:AppB}Calculation of the Improved Effective Potential}

In this Section, we apply the methodology outlined in Section\,\ref{sec:General-Considerations}
to the present theory. We use as a starting point the two-loop effective
potential in Eq. (\ref{eq:61}), from which one can identify the numerical
values of the initial $C_{m,n,p}$ coefficients of the expansion \begin{align}
S_{\mbox{eff}}\left(\phi\right) & =\sum_{n,m,p\geq0}C_{n,m,p}^{\LL}x^{n}y^{m}z^{p}L^{n+m+p-1}\nonumber \\
 & +\sum_{n,m,p\geq0}C_{n,m,p}^{\NLL}x^{n}y^{m}z^{p}L^{n+m+p-2}\nonumber \\
 & +\sum_{n,m,p\geq0}C_{n,m,p}^{\NNLL}x^{n}y^{m}z^{p}L^{n+m+p-3}+\cdots\,,\label{eq:B2}\end{align}
by casting Eq. (\ref{eq:61}) as \begin{align}
S_{\mbox{eff}}^{2\ell}\left(\phi\right)= & y\left(C_{0,1,0}^{\LL}+C_{0,2,0}^{\LL}yL\right)+\left(x^{2}C_{2,1,0}^{\NLL}yL+z^{2}C_{0,1,2}^{\NLL}\right)yL\nonumber \\
 & +\left(C_{4,0,0}^{\NNLL}x^{4}+C_{0,0,4}^{\NNLL}z^{4}+C_{3,0,1}^{\NNLL}x^{3}z+C_{2,0,2}^{\NNLL}x^{2}z^{2}\right)L\,.\label{eq:B3}\end{align}

The beta functions and anomalous dimension that appears in the RGE,\begin{equation}
\left[-\left(1+2\gamma_{\phi}\right)\frac{\partial}{\partial L}+\beta_{y}\frac{\partial}{\partial y}+\beta_{z}\frac{\partial}{\partial z}-6\gamma_{\varphi}\right]S_{\mbox{eff}}\left(\phi\right)=0\,;\label{eq:B4}\end{equation}
were presented in Section\,\ref{sec:appA}, and can be cast as\begin{align}
\beta_{y} & =\beta_{y}^{\left(2\right)}+\beta_{y}^{\left(3\right)}+\beta_{y}^{\left(4\right)}\,,\label{eq:B6}\end{align}
where\begin{align}
\beta_{y}^{\left(2\right)} & =a_{1}y^{2}\,\,;\,\,\beta_{y}^{\left(3\right)}=a_{2}x^{2}y+a_{3}yz^{2}\,,\nonumber \\
\beta_{y}^{\left(4\right)} & =a_{4}x^{4}+a_{5}z^{4}+a_{6}x^{3}z+a_{7}x^{2}z^{2}\,,\label{eq:B7}\end{align}
and\begin{equation}
\beta_{z}=\beta_{z}^{\left(3\right)}=b_{1}z^{3}+b_{2}z^{2}x+b_{3}zx^{2}+b_{4}x^{3}\,,\label{eq:B8}\end{equation}
as for the anomalous dimension, we have \begin{align}
\gamma_{\varphi} & =\gamma_{\varphi}^{\left(2\right)}=c_{1}x^{2}+c_{2}z^{2}\,.\label{eq:B9}\end{align}
The numerical values of the coefficients appearing in the last equations
are\begin{subequations}\label{eq:B10}\begin{align}
a_{1} & =14,\; a_{2}=6c_{1}-\frac{33}{2}=-\left(n^{2}+20\right),\\
a_{3} & =6c_{2}+\frac{15}{4}=4,\, a_{4}=24\left(1+\frac{n^{2}}{8}\right)\,,\\
a_{5} & =-3,\; a_{6}=-6n^{2},\; a_{7}=3n^{2}\,,\\
c_{1} & =-\left(\frac{7}{12}+\frac{n^{2}}{6}\right),\; c_{2}=\frac{1}{24}\,,\\
b_{1} & =-\frac{1}{2},\, b_{2}=\frac{1}{2},\, b_{3}=\frac{97}{72}+\frac{n^{2}}{6},\, b_{4}=\frac{5}{2}\end{align}
\end{subequations}where $n$ is the charge of the fermionic field.

Using these results, we can split Eq.\,(\ref{eq:B4}) according to
the relative powers of coupling constants and logarithms,\begin{align}
 & \left[-\frac{\partial}{\partial L}+\beta_{y}^{\left(2\right)}\frac{\partial}{\partial y}\right]S_{\mbox{eff}}^{\LL}\nonumber \\
 & +\left\{ \left[\beta_{y}^{\left(3\right)}\frac{\partial}{\partial y}+\beta_{z}^{\left(3\right)}\frac{\partial}{\partial z}-6\gamma_{\varphi}^{\left(2\right)}\right]S_{\mbox{eff}}^{\LL}+\left[-\frac{\partial}{\partial L}+\beta_{y}^{\left(2\right)}\frac{\partial}{\partial y}\right]S_{\mbox{eff}}^{\NLL}\right\} \nonumber \\
 & +\left\{ \left[-2\gamma_{x}^{\left(2\right)}\frac{\partial}{\partial L}+\beta_{y}^{\left(4\right)}\frac{\partial}{\partial y}\right]S_{\mbox{eff}}^{\LL}+\left[\beta_{y}^{\left(3\right)}\frac{\partial}{\partial y}+\beta_{z}^{\left(3\right)}\frac{\partial}{\partial z}-6\gamma_{\varphi}^{\left(2\right)}\right]S_{\mbox{eff}}^{\NLL}+\left[-\frac{\partial}{\partial L}+\beta_{y}^{\left(2\right)}\frac{\partial}{\partial y}\right]S_{\mbox{eff}}^{\NNLL}\right\} \nonumber \\
 & =0\label{eq:B11}\end{align}

\subsection{Leading logarithms}

Focusing first on terms of order $x^{m}y^{n}z^{p}L^{n+m+p-2}$ in
Eq.\ (\ref{eq:B11}), one obtains \begin{align}
\left[-\frac{\partial}{\partial L}+\beta_{y}^{\left(2\right)}\frac{\partial}{\partial y}\right]S_{\mbox{eff}}^{\LL}\left(\phi\right) & =\nonumber \\
\left[-\frac{\partial}{\partial L}+a_{1}y^{2}\frac{\partial}{\partial y}\right] & \underset{\left(m+n+p\geq1\right)}{\sum_{m,n,p}}C_{m,n,p}^{\LL}x^{m}y^{n}z^{p}L^{m+n+p-1}=0\label{eq:B12}\end{align}
which furnishes the following relation for the coefficients $C_{n,m,p}^{\LL}$,
\begin{equation}
\left(m+n+p-1\right)C_{m,n,p}^{\LL}-\left(n-1\right)a_{1}C_{m,n-1,p}^{\LL}=0\quad\left(m+n+p\geq2\right)\,.\label{eq:B13}\end{equation}

We find convenient to recast $S_{\mbox{eff}}^{\LL}\left(\phi\right)$
as\begin{align}
S_{\mbox{eff}}^{\LL}\left(\phi\right)= & yS_{\left(0\right)}^{\LL}\left(u\right)+\sum_{m+p\geq1}x^{m}z^{p}L^{m+p-1}S_{\left(m,p\right)}^{\LL}\left(u\right)\,,\label{eq:B14}\end{align}
where\begin{align}
S_{\left(0\right)}^{\LL}\left(u\right)= & \sum_{n\geq0}C_{0,n+1,0}^{\LL}u^{n},\label{eq:B1401}\end{align}
\begin{align}
S_{\left(m,p\right)}^{\LL}\left(u\right)= & \sum_{n\geq0}C_{m,n,p}^{\LL}u^{n},\label{eq:B15}\end{align}
in terms of the variable $u=yL$. Inspection of Eq.\,(\ref{eq:61})
allows one to find the initial coefficient of these sums,\begin{align}
C_{0,1,0}^{LL} & =1,\quad C_{0,2,0}^{LL}=a_{1}\,,\nonumber \\
C_{1,0,0}^{LL} & =C_{0,0,1}^{LL}=C_{1,1,0}^{LL}=C_{1,0,1}^{LL}=C_{0,1,1}^{LL}=C_{2,0,0}^{LL}=C_{0,0,2}^{LL}=0\,.\label{eq:B1301}\end{align}

By looking at Eq.\,(\ref{eq:B13}) with $m=p=0$, we have\begin{equation}
C_{0,n,0}^{\LL}=a_{1}C_{0,n-1,0}^{\LL}\quad\left(n\geq2\right)\,,\label{eq:B16}\end{equation}
with, together with the values $C_{0,1,0}^{LL},C_{0,2,0}^{LL}$ from
Eq.\,(\ref{eq:B1301}), leads to\begin{equation}
C_{0,n,0}^{\LL}=a_{1}^{n-1}\quad\left(n\geq1\right)\label{eq:B17}\end{equation}
hence,\begin{equation}
S_{\left(0\right)}^{\LL}\left(u\right)=\sum_{n\geq0}a_{1}^{n}u^{n}=\frac{1}{1-a_{1}u}\,.\label{eq:B18}\end{equation}

Now setting $m=1$ and $p=0$ in Eq.\,(\ref{eq:B13}), \begin{equation}
nC_{1,n,0}^{\LL}-\left(n-1\right)a_{1}C_{1,n-1,0}^{\LL}=0\quad\left(n\geq1\right)\,,\label{eq:B19}\end{equation}
and from this equation one concludes that $C_{1,1,0}^{\LL}=0$, which
is consistent with the results obtained from the two-loop calculation
of $V_{\mbox{eff}}$ in Eq.\,(\ref{eq:B1301}); this is an important
consistency check of that result. Also from Eq.\,(\ref{eq:B19}),
by recurrence we have\begin{equation}
C_{1,n,0}^{\LL}=0\quad\left(n\geq0\right)\,,\label{eq:B20}\end{equation}
so that $S_{\left(1,0\right)}^{\LL}\left(u\right)=0$. Similar results
are found by setting $m=0$ and $p=1$, i.e., \begin{equation}
C_{0,n,1}^{\LL}=0\quad\left(n\geq0\right)\,,\label{eq:B21}\end{equation}
thus $S_{\left(0,1\right)}^{\LL}\left(u\right)=0$.

Now looking at the terms with $m+p\geq2$ in Eq.\,(\ref{eq:B13}),
for $n=1$ we immediately obtain\begin{equation}
C_{m,1,p}^{\LL}=0\,,\label{eq:B22}\end{equation}
which, by recurrence for larger $n$, implies that\begin{equation}
C_{m,n,p}^{\LL}=0\,.\quad\left(m+p\geq2\right)\label{eq:B23}\end{equation}
Summarizing this results, \begin{equation}
S_{\left(m,p\right)}^{\LL}\left(u\right)=0\quad\left(m,p\neq0\right)\,,\label{eq:B24}\end{equation}
therefore,\begin{equation}
S_{\mbox{eff}}^{\LL}\left(\phi\right)=yS_{\left(0\right)}^{\LL}\left(u\right)=\frac{y}{w}\,,\label{eq:B25}\end{equation}
where we have introduced the definition\begin{equation}
w=1-a_{1}u=1-a_{1}yL\,.\label{eq:B26}\end{equation}

\subsection{Next-to-leading logarithms}

Having found $S_{\mbox{eff}}^{\LL}$, we can now consider terms of
order $x^{m}y^{n}z^{p}L^{m+n+p-3}$ in Eq.\ (\ref{eq:B11}),\begin{equation}
\left[\beta_{y}^{\left(3\right)}\frac{\partial}{\partial y}+\beta_{z}^{\left(3\right)}\frac{\partial}{\partial z}-6\gamma_{\varphi}^{\left(2\right)}\right]S_{\mbox{eff}}^{\LL}+\left[-\frac{\partial}{\partial L}+\beta_{y}^{\left(2\right)}\frac{\partial}{\partial y}\right]S_{\mbox{eff}}^{\NLL}=0\,.\label{eq:B27}\end{equation}
At this point, the first term is completely known, and we proceed
to find out $S_{\mbox{eff}}^{\NLL}$ which, as before, will be written
in the form\begin{align}
S_{\mbox{eff}}^{\NLL}\left(\phi\right)= & y^{2}S_{\left(0\right)}^{\NLL}\left(u\right)+\sum_{m+p\geq1}x^{m}z^{p}L^{m+p-2}S_{\left(m,p\right)}^{\NLL}\left(u\right)\,,\label{eq:B28}\end{align}
\begin{align}
S_{\left(0\right)}^{\NLL}\left(u\right)= & \sum_{n\geq0}C_{m,n+2,p}^{\NLL}u^{n}\,.\label{eq:B2801}\end{align}
\begin{align}
S_{\left(m,p\right)}^{\NLL}\left(u\right)= & \sum_{n\geq0}C_{m,n,p}^{\NLL}u^{n}\,.\label{eq:B29}\end{align}

After some manipulations, Eq.\,(\ref{eq:B27}) can be cast as\begin{align}
\sum_{n\geq1} & \left\{ \left[na_{2}-6c_{1}\right]x^{2}+\left[na_{3}-6c_{2}\right]z^{2}\right\} C_{0,n,0}^{\LL}y^{n}L^{n-1}\nonumber \\
 & +\underset{\left(n\geq1,m+n+p\geq3\right)}{\sum_{m,n,p}}\left[-\left(m+n+p-2\right)C_{m,n,p}^{\NLL}+a_{1}\left(n-1\right)C_{m,n-1,p}^{\NLL}\right]x^{m}y^{n}z^{p}L^{m+n+p-3}\nonumber \\
 & =0.\label{eq:B30}\end{align}
Some initial coefficients for these sums are obtained from Eq.\,(\ref{eq:61}),
as follows,\begin{align}
 & C_{210}^{NLL}=-\frac{33}{2},\quad C_{012}^{NLL}=\frac{15}{4}\,,\nonumber \\
 & C_{011}^{NLL}=C_{020}^{NLL}=C_{002}^{NLL}=C_{110}^{NLL}=C_{101}^{NLL}=C_{200}^{NLL}=0\,,\nonumber \\
 & C_{021}^{NLL}=C_{003}^{NLL}=C_{030}^{NLL}=C_{102}^{NLL}=C_{111}^{NLL}=C_{120}^{NLL}=C_{201}^{NLL}=C_{300}^{NLL}=0\,.\label{eq:B3001}\end{align}

As before, we look at some particular subseries in Eq. (\ref{eq:B30}).
First, isolating terms with $m=2$ and $p=0$,\begin{equation}
\left[na_{2}-6c_{1}\right]C_{0,n,0}^{\LL}-nC_{2,n,0}^{\NLL}+a_{1}\left(n-1\right)C_{2,n-1,0}^{\NLL}=0\,,\label{eq:B33}\end{equation}
which is consistent with the coefficients found in Eq.\,(\ref{eq:B3001}),
since \begin{equation}
\left[a_{2}-6c_{1}\right]C_{0,1,0}^{\LL}-C_{2,1,0}^{\NLL}=6c_{1}-\frac{33}{2}-6c_{1}-\left(-\frac{33}{2}\right)=0\,.\label{eq:B34}\end{equation}
Also from Eq.\,(\ref{eq:B33}), multiplying by $u^{n-1}$ and summing
up over $n$, we obtain a differential equation for the function $S_{\left(2,0\right)}^{\NLL}\left(u\right)$,\begin{equation}
a_{2}u\frac{dS_{\left(0\right)}^{\LL}}{du}+\left(a_{2}-6c_{1}\right)S_{\left(0\right)}^{\LL}-\frac{dS_{\left(2,0\right)}^{\NLL}}{du}+a_{1}u\frac{dS_{\left(2,0\right)}^{\NLL}}{du}=0\,,\label{eq:B35}\end{equation}
or, rewritten in terms of the variable $w=1-a_{1}u$, and according
to Eq.\,(\ref{eq:B25}),\begin{equation}
a_{1}w\frac{dS_{\left(2,0\right)}^{\NLL}}{dw}+\frac{a_{2}}{w^{2}}-\frac{6c_{1}}{w}=0\,.\label{eq:B36}\end{equation}
The solution can be found satisfying the initial condition $S_{\left(2,0\right)}^{\NLL}\left(w=1\right)=0$
as \begin{equation}
S_{\left(2,0\right)}^{\NLL}=\frac{a_{2}}{2a_{1}}\left[\frac{1}{w^{2}}-1\right]-\frac{6c_{1}}{a_{1}}\left[\frac{1}{w}-1\right]\,.\label{eq:B37}\end{equation}

Proceeding similarly for terms with $m=0$ and $p=2$ in Eq.\,(\ref{eq:B30}),
we have\begin{equation}
\left[na_{3}-6c_{2}\right]C_{0,n,0}^{\LL}-nC_{0,n,2}^{\NLL}+a_{1}\left(n-1\right)C_{0,n-1,2}^{\NLL}=0\,,\label{eq:B38}\end{equation}
whose consistency with the initial values in Eq.\,(\ref{eq:B3001})
can also be checked,\begin{equation}
\left[a_{3}-6c_{2}\right]C_{010}^{LL}-C_{012}^{NLL}=6c_{2}+\frac{15}{4}-6c_{2}-\frac{15}{4}=0\,.\label{eq:B39}\end{equation}
Eq.\,(\ref{eq:B38}) furnishes a differential equation for $S_{\left(0,2\right)}^{\NLL}$
whose solution is \begin{equation}
S_{\left(0,2\right)}^{\NLL}=\frac{a_{3}}{2a_{1}}\left[\frac{1}{w^{2}}-1\right]-\frac{6c_{2}}{a_{1}}\left[\frac{1}{w}-1\right]\,.\label{eq:B40}\end{equation}

For all remaining terms in Eq.\,(\ref{eq:B30}), the relation \begin{equation}
-\left(m+n+p-2\right)C_{m,n,p}^{\NLL}+a_{1}\left(n-1\right)C_{m,n-1,p}^{\NLL}=0\,,\label{eq:B41}\end{equation}
together with the initial coefficients $C_{m,0,p}^{\NLL}$, $C_{m,1,p}^{\NLL}$
and $C_{m,2,p}^{\NLL}$ in Eq.\,(\ref{eq:B3001}), implies that $C_{m,n,p}^{\NLL}=0$. 

This way, the only nonvanishing subseries of $S_{\mbox{eff}}^{\NLL}\left(\phi\right)$
are the ones defining $S_{\left(2,0\right)}^{\NLL}$ and $S_{\left(0,2\right)}^{\NLL}$,
and we end up with \begin{equation}
S_{\mbox{eff}}^{\NLL}\left(\phi\right)=x^{2}S_{\left(2,0\right)}^{\NLL}+z^{2}S_{\left(0,2\right)}^{\NLL}\,.\label{eq:B42}\end{equation}

\subsection{Next-to-next to leading logarithms}

Finally, we focus on terms proportional to $x^{m}y^{n}z^{p}L^{m+n+p-4}$,\begin{align}
\left[-2\gamma_{x}^{\left(2\right)}\frac{\partial}{\partial L}+\beta_{y}^{\left(4\right)}\frac{\partial}{\partial y}\right]S_{\mbox{eff}}^{\LL}+\left[\beta_{y}^{\left(3\right)}\frac{\partial}{\partial y}+\beta_{z}^{\left(3\right)}\frac{\partial}{\partial z}-6\gamma_{\varphi}^{\left(2\right)}\right]S_{\mbox{eff}}^{\NLL}\nonumber \\
+\left[-\frac{\partial}{\partial L}+\beta_{y}^{\left(2\right)}\frac{\partial}{\partial y}\right]S_{\mbox{eff}}^{\NNLL} & =0\,.\label{eq:B43}\end{align}
This time we only have information from the two-loop computation of
$V_{\mbox{eff}}$ of the following initial coefficients,\begin{equation}
C_{004}^{N2LL}=-3,\quad C_{202}^{N2LL}=3n^{2},\quad C_{301}^{N2LL}=-6n^{2},\quad C_{400}^{N2LL}=24\left(1+\frac{n^{2}}{8}\right)\,,\label{eq:B44}\end{equation}
so we will focus on the subseries of terms of the form $x^{4}y^{n}L^{n}$,
$z^{4}y^{n}L^{n}$, $x^{2}z^{2}y^{n}L^{n}$, and $x^{3}zy^{n}L^{n}$
in Eq.\,(\ref{eq:B43}).

We start with terms proportional to $x^{4}y^{n}L^{n}$; from Eqs.\,(\ref{eq:B6})
to (\ref{eq:B9}), Eqs.\,(\ref{eq:B25}) and (\ref{eq:B42}), they
arrive from the following terms of Eq.\,(\ref{eq:B43}),\begin{align}
\left[a_{4}x^{4}\frac{\partial}{\partial y}\right]S_{\left(0\right)}^{\LL}+\left[a_{2}x^{2}y\frac{\partial}{\partial y}-6c_{1}x^{2}\right]x^{2}S_{\left(2,0\right)}^{\NLL} & +\left[-\frac{\partial}{\partial L}+a_{1}y^{2}\frac{\partial}{\partial y}\right]S_{\mbox{eff}}^{\NNLL}=0\,,\label{eq:B45}\end{align}
or, writing explicitly, apart from the overall $x^{4}$ factor,\begin{equation}
\sum_{n\geq1}\left[na_{4}C_{0,n,0}^{\LL}u^{n-1}+\left(na_{2}-6c_{1}\right)C_{2,n,0}^{\NLL}u^{n}\right]+\sum_{n\geq0}\left[-\left(n+1\right)C_{4,n,0}^{\NNLL}u^{n}+a_{1}nC_{4,n,0}^{\NNLL}u^{n+1}\right]=0\,.\label{eq:B46}\end{equation}
This relation is consistent with the initial coefficients in Eq.\,(\ref{eq:B45}),
since for the term proportional to $u^{0}$ we have\begin{equation}
a_{4}C_{0,1,0}^{\LL}-C_{4,0,0}^{\NNLL}=a_{4}-24\left(1+\frac{n^{2}}{8}\right)=0\,.\label{eq:B47}\end{equation}
From Eq.\,(\ref{eq:B46}) we obtain the relation \begin{equation}
\left(n+1\right)a_{4}C_{0,n+1,0}^{\LL}u^{n}+\left(na_{2}-6c_{1}\right)C_{2,n,0}^{\NLL}u^{n}-\left(n+1\right)C_{4,n,0}^{\NNLL}u^{n}+a_{1}nC_{4,n,0}^{\NNLL}u^{n+1}=0\quad\left(n\geq1\right)\,,\label{eq:B48}\end{equation}
which provides the following differential equation\begin{align}
a_{4}\left(u\frac{d}{du}+1\right)S_{\left(0\right)}^{\LL}+\left(a_{2}u\frac{d}{du}-6c_{1}\right)S_{\left(2,0\right)}^{\NLL} & +\left(u\left(a_{1}u-1\right)\frac{d}{du}-1\right)S_{\left(4,0\right)}^{\NNLL}=0\label{eq:B49}\end{align}
to be solved for\begin{equation}
S_{\left(4,0\right)}^{\NNLL}\left(u\right)=\sum_{n=0}C_{4,n,0}^{\NNLL}u^{n}\,.\label{eq:B50}\end{equation}
Eq.\,(\ref{eq:B49}) is more easily solved when written in terms
of the variable $w=1-a_{1}yL$,\begin{align}
\left(w\left(w-1\right)\frac{d}{dw}+1\right)S_{\left(4,0\right)}^{\NNLL}= & \left(a_{2}\left(w-1\right)\frac{d}{dw}-6c_{1}\right)S_{\left(2,0\right)}^{\NLL}\nonumber \\
 & +a_{4}\left(\left(w-1\right)\frac{d}{dw}+1\right)S_{\left(0\right)}^{\LL}\,.\label{eq:B5001}\end{align}
The solution $S_{\left(4,0\right)}^{\NNLL}$ is \begin{equation}
S_{\left(4,0\right)}^{\NNLL}=\frac{\alpha_{3}}{w^{3}}+\frac{\alpha_{2}}{w^{2}}+\frac{\alpha_{1}}{w}+\alpha_{0}\,,\label{eq:B51}\end{equation}
where the coefficients $\alpha_{i}$ are\begin{subequations}\label{eq:B52}\begin{align}
\alpha_{3} & =\frac{a_{2}^{2}}{4a_{1}}\,,\\
\alpha_{2} & =-\frac{3a_{2}c_{1}}{a_{1}}-\frac{a_{2}^{2}}{12a_{1}}+\frac{a_{4}}{3}\,,\\
\alpha_{1} & =\frac{18c_{1}^{2}}{a_{1}}-\frac{a_{2}^{2}}{12a_{1}}+\frac{a_{4}}{3}\,,\\
\alpha_{0} & =\frac{3a_{2}c_{1}}{a_{1}}-\frac{18c_{1}^{2}}{a_{1}}-\frac{a_{2}^{2}}{12a_{1}}+\frac{a_{4}}{3}\,.\end{align}
\end{subequations}

Proceeding similarly for terms of the form $z^{4}y^{n}L^{n}$, we
obtain the relation\begin{align}
\left(n+1\right)a_{5}C_{0,n+1,0}^{\LL}u^{n}+\left(na_{3}+2b_{1}-6c_{2}\right)C_{0,n,2}^{\NLL}u^{n}\nonumber \\
-\left(n+1\right)C_{0,n,4}^{\NNLL}u^{n}+a_{1}nC_{0,n,4}^{\NNLL}u^{n+1} & =0\quad\left(n\geq0\right)\,,\label{eq:B53}\end{align}
which provides us a differential equation for the determination of\begin{equation}
S_{\left(0,4\right)}^{\NNLL}=\sum_{n=0}C_{0,n,4}^{\NNLL}u^{n}\,,\label{eq:B55}\end{equation}
as follows,\begin{align}
a_{5}\left(u\frac{d}{du}+1\right)S_{\left(0\right)}^{\LL}+\left(a_{3}u\frac{d}{du}+2b_{1}-6c_{2}\right)S_{\left(0,2\right)}^{\NLL}+\left(u\left(a_{1}u-1\right)\frac{d}{du}-1\right)S_{\left(0,4\right)}^{\NNLL} & =0\,.\label{eq:B56}\end{align}
The solution, again in terms of the variable $w$, is\begin{equation}
S_{\left(0,4\right)}^{\NNLL}=\frac{\beta_{3}}{w^{3}}+\frac{\beta_{2}}{w^{2}}+\frac{\beta_{1}}{w}+\beta_{0}\,,\label{eq:B57}\end{equation}
where\begin{subequations}\label{eq:B58}\begin{align}
\beta_{3} & =\frac{a_{3}^{2}}{4a_{1}}\,,\\
\beta_{2} & =\frac{a_{3}b_{1}}{3a_{1}}-\frac{3a_{3}c_{2}}{a_{1}}-\frac{a_{3}^{2}}{12a_{1}}+\frac{a_{5}}{3}\,,\\
\beta_{1} & =-\frac{6b_{1}c_{2}}{a_{1}}+\frac{a_{3}b_{1}}{3a_{1}}+\frac{18c_{2}^{2}}{a_{1}}-\frac{a_{3}^{2}}{12a_{1}}+\frac{a_{5}}{3}\,,\\
\beta_{0} & =\frac{6b_{1}c_{2}}{a_{1}}-\frac{2a_{3}b_{1}}{3a_{1}}+\frac{3a_{3}c_{2}}{a_{1}}-\frac{18c_{2}^{2}}{a_{1}}-\frac{a_{3}^{2}}{12a_{1}}+\frac{a_{5}}{3}\,.\end{align}
\end{subequations}

Now, focusing on terms proportional to $x^{2}z^{2}y^{n}L^{n}$, we
obtain the relation\begin{align}
\left(n+1\right)a_{7}C_{0,n+1,0}^{\LL}u^{n}+\left(a_{2}C_{0,n,2}^{\NLL}+a_{3}C_{2,n,0}^{\NLL}\right)nu^{n}+\left(2b_{3}-6c_{1}\right)C_{0,n,2}^{\NLL}u^{n}\nonumber \\
-6c_{2}C_{2,n,0}^{\NLL}u^{n}-\left(n+1\right)C_{2,n,2}^{\NNLL}u^{n}+a_{1}nC_{2,n,2}^{\NNLL}u^{n+1} & =0\,.\label{eq:B59}\end{align}
The function \begin{equation}
S_{\left(2,2\right)}^{\NNLL}=\sum_{n=0}C_{2,n,2}^{\NNLL}u^{n}\,,\label{eq:B61}\end{equation}
is determined by the equation\begin{align}
a_{7}\left(u\frac{dS_{\left(0\right)}^{\LL}}{du}+S_{\left(0\right)}^{\LL}\right)+\left(a_{2}u\frac{d}{du}+2b_{3}-6c_{1}\right)S_{\left(0,2\right)}^{\NLL}\nonumber \\
+\left(a_{3}u\frac{d}{du}-6c_{2}\right)S_{\left(2,0\right)}^{\NLL}+\left(u\left(a_{1}u-1\right)\frac{d}{du}-1\right)S_{\left(2,2\right)}^{\NNLL} & =0\,.\label{eq:B62}\end{align}
whose solution is\begin{equation}
S_{\left(2,2\right)}^{\NNLL}=\frac{\gamma_{3}}{w^{3}}+\frac{\gamma_{2}}{w^{2}}+\frac{\gamma_{1}}{w}+\gamma_{0}\,,\label{eq:B63}\end{equation}
where\begin{subequations}\label{eq:B6301}\begin{align}
\gamma_{3} & =\frac{a_{2}a_{3}}{2a_{1}}\,,\\
\gamma_{2} & =\frac{a_{3}b_{3}}{3a_{1}}-\frac{3a_{3}c_{1}}{a_{1}}-\frac{3a_{2}c_{2}}{a_{1}}-\frac{a_{2}a_{3}}{6a_{1}}+\frac{a_{7}}{3}\,,\\
\gamma_{1} & =-\frac{6b_{3}c_{2}}{a_{1}}+\frac{a_{3}b_{3}}{3a_{1}}+\frac{36c_{1}c_{2}}{a_{1}}-\frac{a_{2}a_{3}}{6a_{1}}+\frac{a_{7}}{3}\,,\\
\gamma_{0} & =\frac{6b_{3}c_{2}}{a_{1}}-\frac{2a_{3}b_{3}}{3a_{1}}+\frac{3a_{3}c_{1}}{a_{1}}+\frac{3a_{2}c_{2}}{a_{1}}-\frac{36c_{1}c_{2}}{a_{1}}-\frac{a_{2}a_{3}}{6a_{1}}+\frac{a_{7}}{3}\,.\end{align}
\end{subequations}

Finally, summing up terms of the form $x^{3}zy^{n}L^{n}$, we have
the relation\begin{equation}
\left(n+1\right)a_{6}C_{0,n+1,0}^{\LL}u^{n}+2b_{4}C_{0,n,2}^{\NLL}u^{n}-\left(n+1\right)C_{3,n,1}^{\NNLL}u^{n}+a_{1}nC_{3,n,1}^{\NNLL}u^{n+1}=0\,,\label{eq:B64}\end{equation}
which determines \begin{equation}
S_{\left(3,1\right)}^{\NNLL}=\sum_{n\geq0}C_{3,n,1}^{\NNLL}u^{n}\,,\label{eq:B66}\end{equation}
by the equation\begin{equation}
a_{6}\left(u\frac{d}{du}+1\right)S_{\left(0\right)}^{\LL}+2b_{4}S_{\left(0,2\right)}^{\NLL}+\left(u\left(a_{1}u-1\right)\frac{d}{du}-1\right)S_{\left(3,1\right)}^{\NNLL}=0\,.\label{eq:B67}\end{equation}
The solution reads \begin{equation}
S_{\left(3,1\right)}^{\NNLL}=\frac{\delta_{2}}{w^{2}}+\frac{\delta_{1}}{w}+\delta_{0}\,,\label{eq:B68}\end{equation}
with\begin{subequations}\label{eq:B69}\begin{align}
\delta_{2} & =\frac{a_{3}b_{4}}{3a_{1}}+\frac{a_{6}}{3}\,,\\
\delta_{1} & =-\frac{6b_{4}c_{2}}{a_{1}}+\frac{a_{3}b_{4}}{3a_{1}}+\frac{a_{6}}{3}\,,\\
\delta_{0} & =\frac{6b_{4}c_{2}}{a_{1}}-\frac{2a_{3}b_{4}}{3a_{1}}+\frac{a_{6}}{3}\,.\end{align}
\end{subequations}

As a result, \begin{align}
S_{\mbox{eff}}^{\NNLL} & =\left(x^{4}S_{\left(4,0\right)}^{\NNLL}+z^{4}S_{\left(0,4\right)}^{\NNLL}+x^{2}z^{2}S_{\left(2,2\right)}^{\NNLL}+x^{3}zS_{\left(3,1\right)}^{\NNLL}\right)L\,.\label{eq:B70}\end{align}

\section{\label{sec:Dynamical-Breaking}Dynamical Breaking of Symmetry}

In this section, we show how the dynamical breaking of conformal symmetry
occurs in the present theory, taking into account the improved effective
potential we have obtained, \begin{align}
V_{\mbox{eff}}\left(\phi\right)= & \frac{\pi}{6}\phi^{6}\left\{ S_{\mbox{eff}}^{\LL}\left(\phi\right)+S_{\mbox{eff}}^{\NLL}\left(\phi\right)+S_{\mbox{eff}}^{\NNLL}\left(\phi\right)+\kappa\right\} ,\label{eq:14}\end{align}
$\kappa$ being a finite renormalization constant, which is determined
by imposing the tree level definition of the coupling constant \begin{equation}
\left.\frac{d^{6}V_{\mbox{eff}}\left(\phi\right)}{d^{6}\phi}\right|_{\phi^{2}=\mu}=\frac{d^{6}V_{\mbox{tree}}\left(\phi\right)}{d\phi^{6}}=6!\pi^{2}y\,.\label{eq:15}\end{equation}
The fact that $V_{\mbox{eff}}\left(\phi\right)$ has a minimum at
$\phi^{2}=\mu$ requires that \begin{equation}
\left.\frac{dV_{\mbox{eff}}\left(\phi\right)}{d\phi}\right|_{\phi^{2}=\mu}=0\,,\label{eq:16}\end{equation}
and this equation is used to determine the value of $y$ as a function
of the free parameters $x$, $z$ and $n$. This give us a seventh-degree
equation in $y$, and among its solutions we will look for those which
are real and positive, and correspond to a minimum of the potential,
i.e., \begin{equation}
m_{\varphi}^{2}=\left.\frac{d^{2}V_{\mbox{eff}}\left(\phi\right)}{d\phi^{2}}\right|_{\phi^{2}=\mu}>0\,.\label{eq:17}\end{equation}

\begin{figure}
\begin{centering}
\includegraphics[width=16cm]{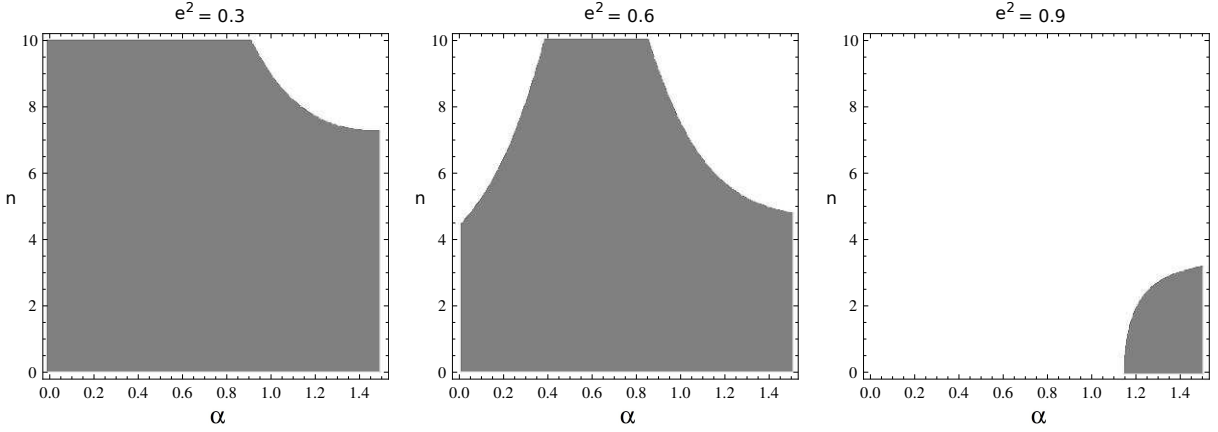}
\par\end{centering}

\caption{\label{fig:ex01}Sections of the parameter space of constant $e^{2}$,
showing where the dynamical symmetry breaking occurs, using the improved
effective potential. }

\end{figure}

\begin{figure}
\begin{centering}
\includegraphics[width=15cm]{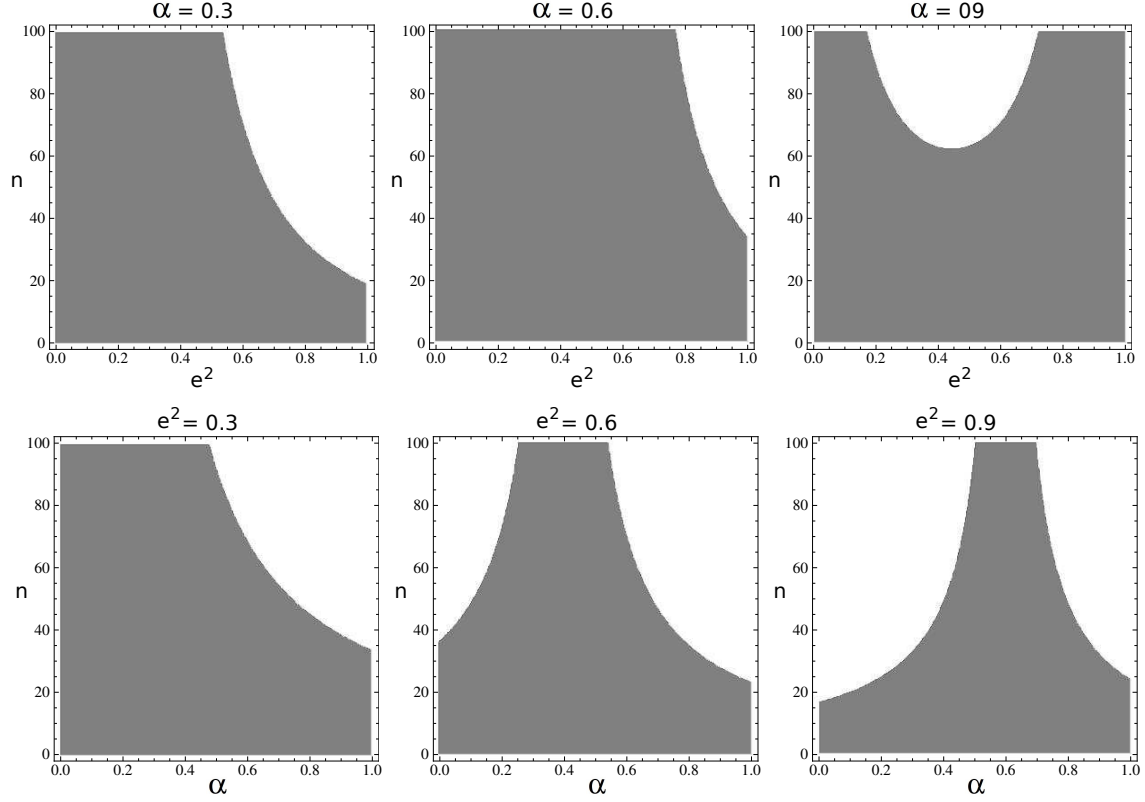}
\par\end{centering}

\caption{\label{fig:unimproved}Sections of the parameter space of constant
$e^{2}$ or $\alpha$, showing where the dynamical symmetry breaking
occurs, using the unimproved 2-loop calculation of the effective potential. }

\end{figure}

\begin{figure}
\begin{centering}
\includegraphics[width=15cm]{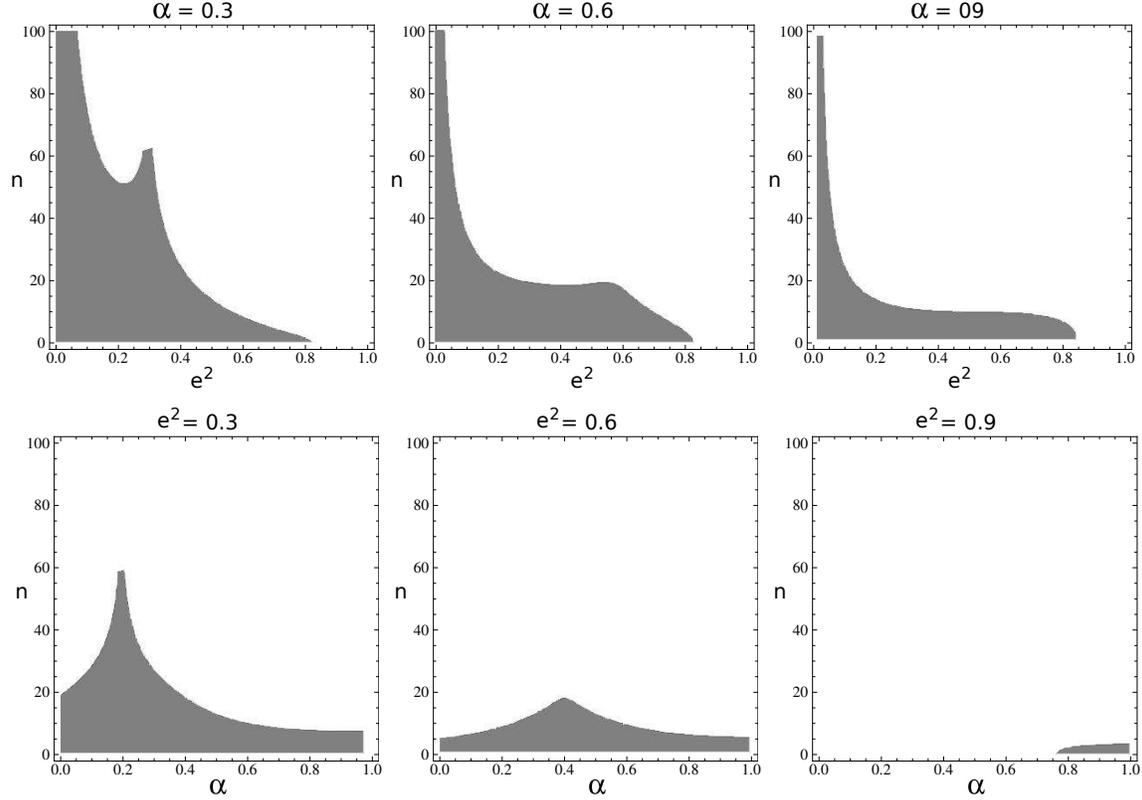}
\par\end{centering}

\caption{\label{fig:improved}Same as Fig.\,\ref{fig:unimproved}, but using
the improved effective potential. It is apparent that when $e^{2}=\alpha$,
the effective potential is stable for higher values of $n$; this
feature can also be seen in Fig.\,\ref{fig:unimproved}.}

\end{figure}

We explore the parameter space of the constants $x$, $z$, $n$,
looking for values where the dynamical symmetry breaking is operational
at the perturbative level. This can be done either using the unimproved
effective potential in Eqs. (\ref{eq:5},\ref{eq:61}), or the improved
one in Eq. (\ref{eq:14}). This latter yields much stronger constraints
on the parameter space of the theory, thus providing a much finer
inspection on the dynamical breaking of the conformal symmetry in
this model. This fact becomes manifest if we plot sections of the
parameter space highlighting the region where a valid $y$ could be
found. Plots for $e^{2}=0.3$, $0.6$, and $0.9$ are shown in Fig.\,\ref{fig:ex01};
for the same range of the parameters, the unimproved effective potential
would pose no restrictions. As an example, for $e^{2}=0.9$ and $n=1$,
from Fig. \ref{fig:ex01} we obtain the restriction $\alpha>1.15$,
so in principle a lower bound $7.99975\mu^{2}$ for the mass of the
scalar is predicted. No such prediction could be made, in this case,
using the unimproved effective potential. For larger $n$, this effect
is still more dramatic: in Figs.\,\ref{fig:unimproved} and\,\ref{fig:improved}
we plot several sections of the parameter space, considering the unimproved
and the improved effective potentials, respectively.

Another interesting fact is that, for certain values of $x$, $z$,
and $n$, Eq. (\ref{eq:17}) provides \emph{two} viable solutions
for $y$. This is true both for the unimproved as well as for the
improved effective potential. For example, for $e^{2}=0.5$, $\alpha=0.5$
and $n=1$, the unimproved potential leads to the equation \begin{equation}
-200.852y^{2}+60.376y-0.0120817=0\,,\label{eq:18}\end{equation}
for the determination of $y$, from which we obtain two solutions\begin{subequations}\label{eq:19}\begin{align}
y_{1} & =0.30039\,,\\
y_{2} & =0.00020\,.\end{align}
\end{subequations}The corresponding masses predicted for the scalar
are $m_{1}=7.7907\mu^{2}$ and $m_{2}=0.00519\mu^{2}$. For the same
value values of the parameters $e^{2}$, $\alpha$ and $n$, the improved
effective potential yields\begin{align}
-4.75607\times10^{9}y^{7}-4.75649\times10^{8}y^{6}-2.46246\times10^{7}y^{5}-882982.y^{4}\nonumber \\
-24137.9y^{3}-471.335y^{2}+60.0824y-0.0379559 & =0\,,\label{eq:20}\end{align}
whose positive and real solutions are\begin{subequations}\label{eq:21}\begin{align}
y_{1} & =0.02540\,,\\
y_{2} & =0.00063\,,\end{align}
\end{subequations}providing $m_{1}=0.18595\mu^{2}$ and $m_{1}=0.015843\mu^{2}$.

\begin{figure}
\begin{centering}
\includegraphics[width=10cm]{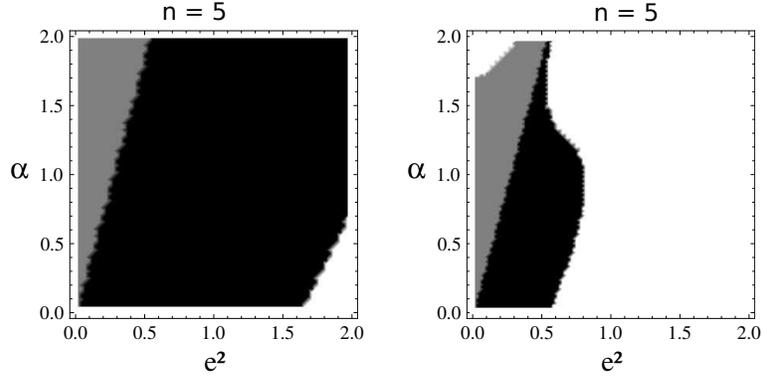}
\par\end{centering}

\caption{\label{fig:twosols01}Regions of the $e^{2}$-$\alpha$ plane, for
$n=1$, painted according to the number of viable solutions $y$ for
Eq. (\ref{eq:17}) for the unimproved effective potential (left) and
for the improved one (right). Black, gray and white means two, one,
and none solutions, respectively.}

\end{figure}

\begin{figure}
\begin{centering}
\includegraphics[width=15cm]{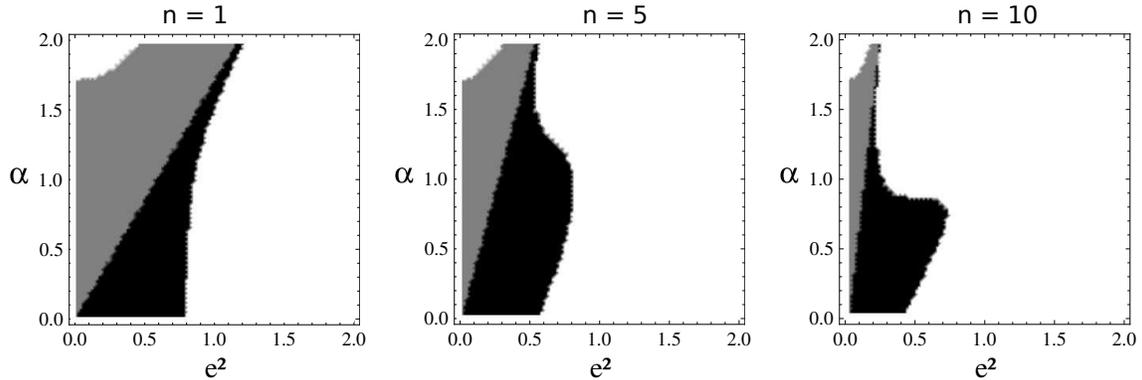}
\par\end{centering}

\caption{\label{fig:twosols02}Same as Fig. \ref{fig:twosols01} (right), but
for different values of $n$. For larger $n$, the region where we
found a unique solution for the conformal symmetry breaking becomes
smaller in absolute terms, and also in comparison to the region where
we found two solutions.}

\end{figure}

Figure\,\ref{fig:twosols01} depicts the region of the $\alpha-e^{2}$
plane, for $n=5$, where such a duplicity of solutions occurs, both
for the unimproved and improved effective potentials. The most important
difference between the two cases is that the improved effective potential
drastically reduces the range of parameters where the duplicity happens.
Figure\,\ref{fig:twosols02} shows how the situation changes for
different values of $n$, for the second case. 

\begin{figure}
\begin{centering}
\includegraphics[width=15cm]{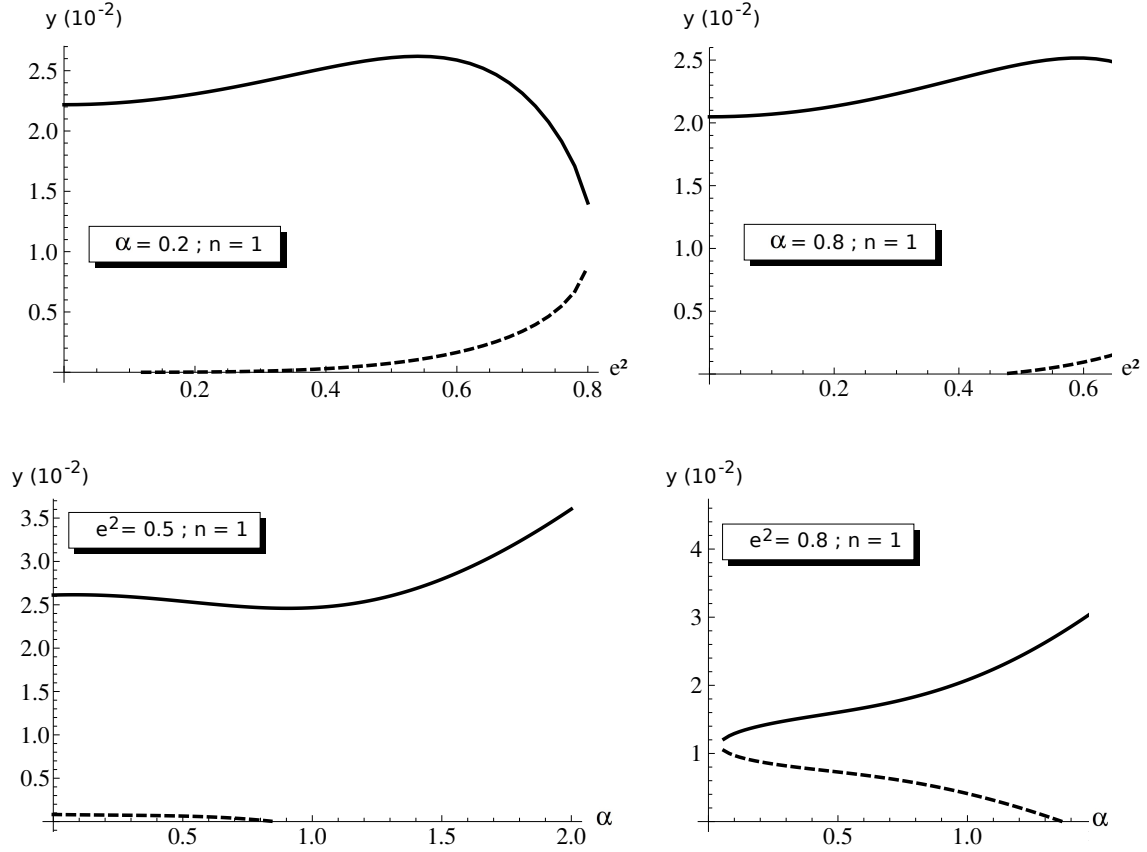}
\par\end{centering}

\caption{\label{fig:twosols03}Behavior of the two solutions $y_{1}$ and $y_{2}$
(solid and dashed lines, respectively) when varying the parameters
of the model. }

\end{figure}

The pattern in Eqs. (\ref{eq:20},\ref{eq:21}) is quite typical:
the solution $y_{2}$ is smaller than $y_{1}$. Fixing the parameters
$e^{2}=0.5$ and $n=1$, $y_{2}$ becomes smaller as $\alpha$ increases.
At some point, the solution $y_{1}$ approaches zero and becomes negative,
so it is not counted anymore as a viable solution. This behavior is
clearly visible at the first graph in Fig. \ref{fig:twosols03}. For
fixed $\alpha$ and $n$, the situation is reversed: $y_{2}$ becomes
smaller as $\alpha$ decreases, as also seen in Fig. \ref{fig:twosols03}.

In summary, there are regions of the parameter space of the theory
where there are \emph{two }possible vacua, in which the conformal
symmetry was broken by radiative corrections. The scalar selfcoupling
and mass are clearly different for these two vacua. Our numerical
studies show, however, that for the improved effective potential,
the region of the parameter space where such a situations takes place
is much smaller than for the unimproved potential.

\section{\label{sec:Conclusions}Conclusions}

The Renormalization Group Equation is well known to provide better
approximations to the effective potential of a given model than a
pure perturbative calculation up to a given loop order. In this work,
we pursued the idea of using the RGE to sum infinite subseries of
the expansion of the effective potential in powers of coupling constants
and logarithms $L=\ln\left(\phi^{2}/\mu\right)$. 

We focused on a Chern-Simons theory coupled to a fermion and a complex
scalar field. Renormalization group beta-functions and anomalous dimensions
should be known up to the two-loop order; we collected results already
available in the literature and calculated the beta-function for the
Yukawa coupling and the wavefunction renormalization of the fermionic
field. With this information, we were able to use the RGE to extract
the maximum amount of information of the perturbative calculation,
obtaining and improved effective potential which, in principle, should
allows us to establish more precisely the properties of the model.
In particular, we were interested in studying the phase where the
conformal symmetry breaking of the model is broken by the radiative
corrections. 

By comparing the outcomes of the standard analysis of dynamical symmetry
breaking in the model using the standard effective action calculated
from loop corrections and the improved one, we shown how the latter
indeed provides a more precise determination of the properties of
the model in the broken phase. This should serve as an instructive
example of the relevance of using the RGE to obtain the maximum amount
of information on the effective action from a given perturbative calculation.
This idea is quite relevant in the context of models with classical
conformal invariance which is broken at the quantum level, for the
sake of obtaining the most precise predictions. 

It would be interesting to extend the calculations discussed in this
work to higher loop orders, to see whether this would imply in some
mild refinement of the results presented here, or some even more drastic
reduction of the parameter space region where the dynamical symmetry
breaking happens.

\vspace{1cm}

\textbf{Acknowledgments. }The authors thank M. Gomes for reading the
manuscript. This work was partially supported by the Brazilian agencies
Conselho Nacional de Desenvolvimento Cient\'{\i}fico e Tecnol\'{o}gico
(CNPq) and Funda\c{c}\~{a}o de Amparo \`{a} Pesquisa do Estado de
S\~{a}o Paulo (FAPESP).

\end{document}